\documentclass[12pt]{article}

\usepackage{epsf}
\usepackage{epsfig}
\usepackage{cite}
\usepackage{amsmath}
\usepackage{graphics}

\begin{document}

\setlength{\textheight}{21.5cm}
\setlength{\oddsidemargin}{0.cm}
\setlength{\evensidemargin}{0.cm}
\setlength{\topmargin}{0.cm}
\setlength{\footskip}{1cm}
\setlength{\arraycolsep}{2pt}

\renewcommand{\thefootnote}{\#\arabic{footnote}}
\setcounter{footnote}{0}

\newcommand{\gtrsim}{ \mathop{}_{\textstyle \sim}^{\textstyle >} }
\newcommand{\lesssim}{ \mathop{}_{\textstyle \sim}^{\textstyle <} }
\newcommand{\rem}[1]{{\bf #1}}
\renewcommand{\thefootnote}{\fnsymbol{footnote}}
\setcounter{footnote}{0}
\def\thefootnote{\fnsymbol{footnote}}

\hfill {\tt arXiv:0806.1707 [gr-qc]}\\

\vskip .5in

\begin{center}

\bigskip
\bigskip

{\Large \bf High Longevity Microlensing Events and Dark Matter Black Holes}

\vskip .45in

{\bf Paul H. Frampton} 

\vskip .3in

{\it Department of Physics and Astronomy, University of North Carolina,
Chapel Hill, NC 27599-3255.}

\end{center}

\vskip .4in 
\begin{abstract}
Gravitational microlensing has been employed to identify massive halo objects 
by their amplilification of distant sources; MACHO 
searches have studied event times   
$2 h \lesssim t_0 \lesssim 2 y$ corresponding
masses in the range $10^{-6} M_{\odot} \lesssim M \lesssim 100 M_{\odot}$.
We suggest that larger masses up to $10^6 M_{\odot}$
are also of considerable interest.
It has not been excluded that there is a significant number of halo
black holes with such high masses as suggested by 
cosmological entropy considerations and potentially
detectable by high longevity microlensing events.
\end{abstract}

\renewcommand{\thepage}{\arabic{page}}
\setcounter{page}{1}
\renewcommand{\thefootnote}{\#\arabic{footnote}}

\newpage

\noindent {\it Introduction}. ~~ 

\bigskip

Attempts to unify gravity with the other interactions of Nature
may be guided by the holographic\cite{holo} principle which
provides an upper limit on the amount of information
which can be contained is a given three-dimensional volume in
terms of its two-dimensional surface area. Although the principle
is not proven rigorously, it provides the best available
guide to estimates of cosmological entropy and hence to suggest
which future observations can help to quantify where the entropy
lies. It may well be objects not yet detected for want of a
motivation to make the requisite observations. One such set of
observations is the subject of the present article.

\bigskip

Our present aim is {\it not} to explain all or even most of the dark
matter which could be {\it e.g.} WIMPS but rather to suggest
that a small fraction (say, 1\%) may account for a large
fraction (say, 90\%) of the entropy. Thus, if a ``pie chart''
were drawn for entropy, rather than energy, the appearance
would be dramatically different.

\bigskip

Here we assume throughout the holographic principle is physically 
correct that the upper limit of entropy, taking into account all
degrees of freedom, both gravitational and non-gravitational,
is the area of the surface surrounding a volume in units of the
Planck area. Taking a standard estimate for the volume and area of the
visible universe, this gives an upper limit for the entropy of
the universe. The value is $(S_U)^{max} \sim 10^{123}$.

\bigskip

The conventional wisdom (see {\it e.g.} \cite{DMBH})
is that the present entropy of the universe
is overwhelmingly dominated by the supermassive black holes (SMBHs) at the
cores of galaxies. This provides a lower limit on the cosmological entropy
which, taking for simplicity $10^{12}$ galaxies each containing a
SMBH of mass $10^{7} M_{\odot}$, gives $(S_U)^{min} \sim 10^{103}$
since the entropy of a black hole with $M_{BH} = \eta M_{\odot}$
is $S_{BH} (\eta M_{\odot}) \sim 10^{77} \eta^2$.
If we further acknowledge that the galaxies are receding from one another
at an accelerated rate such that coalescence is, in general, unlikely
and they can be regarded as totally segregated and disentangled then
the upper limit on $S_U$ is refined to $(S_U)^{max} \sim 10^{111}$.
This diminution from $10^{123}$ to $10^{111}$ arises from dividing
out the number $10^{12}$ of galaxies since the maximum total entropy
of the universe becomes the sum of the maximum possible entropies
of the separate galaxies.

\bigskip

This provides 
the cosmological entropy range\footnote{These limits apply to
the visible universe not to a single galaxy.}
\begin{equation}
103 \lesssim \log_{10} S_U \lesssim 111
\label{entropy}
\end{equation}
the first of two interesting
windows 
which are the subject for this Letter. Conventional wisdom is
$S_U \sim (S_U)^{min} = 10^{103}$.

\newpage

\bigskip
\bigskip

\noindent {\it Dark Matter Black Holes}

\bigskip

If we consider normal baryonic matter, other than black holes,
contributions to the entropy are far smaller. The background
radiation and relic neutrinos each provide $\sim 10^{88}$.
We have learned in the last decade about the dark side
of the universe. WMAP\cite{WMAP} suggests that the pie slices
for the overall energy include 24\% dark matter
and 72\% dark energy. Dark energy has no known microstructure,
and especially if it is characterized only by a cosmological
constant, may be assumed to have zero entropy. As already
mentioned, the baryonic matter other than the SMBHs contributes
far less than $(S_U)^{min}$.

\bigskip

This leaves the dark matter which is concentrated in halos
of galaxies and clusters.

\bigskip

It is counter to the second law of thermodynamics, if
higher entropy configurations are available,
that essentially all the entropy
of the universe is concentrated in only the known
supermassive black holes (SMBH). The Schwarzschild
radius for a $10^7 M_{\odot}$ SMBH is $\sim 3 \times 10^7$ km
and so $10^{12}$ of them occupy only $\sim 10^{-36}$
of the volume of the visible universe
\footnote{With dark matter black holes (DMBH) this fraction
is a few times $10^{-35}$ so the present proposal 
makes only a tiny change to the surprising compression
of the total entropy but does suggest what can
constitute a far bigger fraction of entropy
than the SMBHs.}. 

\bigskip

Several years ago important work by Xu and Ostriker\cite{Ostriker}
showed by numerical simulations that DMBHs with masses above
$10^6 M_{\odot}$ would have the property of disrupting
the dynamics of a galactic halo leading to 
runaway spiral into the center. This provides
an upper limit $(M_{DMBH})^{max} \sim 10^6 M_{\odot}$.

\bigskip

Gravitational lensing observations are amongst the most useful
for determining the mass distributions of dark matter. Weak
lensing by, for example, the HST shows the strong distortion
of radiation from more distant galaxies by the mass of the
dark matter and leads to astonishing three-dimensional maps
of the dark matter trapped within clusters. At the scales
we consider $\sim 3 \times 10^7$ km, however, weak lensing
has no realistic possibility of detecting DMBHs in the
forseeable future.

\bigskip

Gravitational microlensing presents a much more optimistic
possibility. This technique which exploits the amplification
of a distant source was first emphasized in modern times
(Einstein considered it in 1912 unpublished
work) by Paczynski\cite{Paczynski}. Subsequent observations
\cite{Alcock,rich} found many examples of MACHOs, yet insufficient
to account for all of the halo by an order of magnitude.
These MACHO searches looked for masses in the range
$10^{-6} M_{\odot}
\lesssim M \lesssim 10^2 M_{\odot}$.

\bigskip

According to \cite{Paczynski} the time $t_0$ of a microlensing
event is given by

\begin{equation}
t_0 \equiv \frac{r_E}{v}
\label{t0}
\end{equation}
where $r_E$ is the Einstein radius and $v$ is the
lens velocity usually taken as $v = 200$ km/s. The radius
$r_E$ is proportional to the square root of the lens mass
and numerically one finds
\begin{equation}
t_0 \simeq 0.2 y ~ \left( \frac{M}{M_{\odot}} \right)^{1/2}
\label{t0value}
\end{equation}
so that,  for the MACHO masses considered, $2 h \lesssim t_0
\lesssim 2 y$. Although some of the already observed
MACHOs may be DMBHs, they do not saturate the possible
mass or entropy for dark matter so let us
set as definition $(M_{DMBH})^{min} \sim 10^2 M_{\odot}$.
This provides the range for DMBH mass
\begin{equation}
2 \lesssim \log_{10} \eta = \log_{10} (M_{DMBH}/M_{\odot})
\lesssim 6
\label{DMBHmass}
\end{equation}
which, after Eq.(\ref{entropy}), provides a second window
of interest. It corresponds to
$2 y \lesssim t_0 \lesssim 200 y$. Ranges (\ref{entropy}) 
and (\ref{DMBHmass})
are related in the next section.

\newpage

\bigskip
\bigskip

\noindent {\it Cosmological entropy considerations}

\bigskip

As mentioned already, the key guide will be the
holographic principle\cite{holo} which informs us
that the cosmological entropy is in the window
(\ref{entropy}). It cannot be at the absolute maximum
value because that is possible only if every halo
has already completely collpsed into a single
black hole.

\bigskip

Also, the absolute minimum although not excluded seems
intuitively implausible because all the entropy
is compressed into $10^{-36} V_U$. 

\bigskip

The natural suggestion is that there exist DMBHs
in the mass region (\ref{DMBHmass}). The number is
limited by the total halo mass $10^{12} M_{\odot}$.
The total entropy is higher for higher DMBH mass
because $S \propto M^2$. Let $n$ be the number of
DMBHs per halo, $\eta$ be the ratio ($ M_{DMBH}/M_{\odot}$), 
$S_U$ be the total entropy for $10^{12}$ halos and
$t_0$ be the microlensing longevity.
The Table shows five possibilities. Each corresponds to
the dark matter black holes
contributing an average density $\rho_{DMBH} \sim 1\% \rho_{DM}$
where $\rho_{DM}$ is the mean dark matter density.
This is to be compared to the average density
contributed by the known supermassive black holes
$\rho_{SMBH} \sim 0.001\% \rho_{DM}$.

\begin{center}

{\bf Dark Matter Black Holes and Microlensing Longevity}

\bigskip
\bigskip

\begin{tabular}{||c|c|c|c||c||}
\hline\hline
$\log_{10} n$  & $ \log_{10} \eta$  & $ \log_{10} S_{halo}$ & $\log_{10}
S_U$ & $t_0$ (years) \\
 \hline\hline
 8 & 2 & 88 & 100 & 2   \\
 \hline
 7 & 3 & 89 & 101 & 6 \\
 \hline
6  & 4 & 90 & 102 & 20 \\
 \hline
5 & 5 & 91 & 103 & 60 \\
\hline
4 & 6 & 92 & 104 & 200 \\
\hline\hline
 \end{tabular} 

  \end{center} 

  \bigskip 

\newpage

{\it Observation of Dark Matter Black Holes}

\bigskip

Since microlensing observations \cite{Alcock,rich}
already impinge on the lower end of the range
(\ref{DMBHmass}) and the Table, it is likely that observations
which look at longer time periods, have
higher statistics or sensitivity to the
period of maximum
amplification can detect heavier
mass DMBHs in the halo.

\bigskip

If this can be achieved, and it seems a worthwhile
enterprise, then the known entropy of the universe
could be increased by an order of magnitude. There exists an interesting
analysis\cite{Yoo} of wide binaries which places a weak limit
on DMBHs which perhaps can be strengthened? To my knowledge, the DMBHs
as listed in the Table, and contributing $\sim 1\% \rho_{DM}$
are not excluded by existing observations.

\bigskip

Previous analysis\cite{Yoo,Murali,Moore} have assigned upper limits
on the fraction ($f$) of the halo mass that can be constituted by DMBHs.
These include $f < 0.5$ \cite{Yoo} and $f<0.025$ \cite{Murali}.
We have no reason to suggest that all of the dark matter
halo mass is from
DMBHs so the fraction $f$ could indeed be very small.
Yet DMBHs can still provide a very large fraction of the  entropy of
the universe. 
For example, taking\cite{Murali} $f = 0.01$ and $10^6 M_{\odot}$ as 
mass allows $10^4$ DMBHs per halo,
a total of $\sim 10^{16}$ Mega-$M_{\odot}$ black holes
in the universe and the 
fraction of the total entropy of the universe
provided by dark matter black holes is $\sim 90\%$.

\bigskip

It is this entropy argument based on holography and the second law of
thermodynamics which is the most
compelling supportive argument for DMBHs.
If each galaxy halo asymptotes to a black hole the final entropy 
of the universe will be $\sim 10^{111}$
as in Eq.(\ref{entropy}) and the universe will
contain just $\sim 10^{12}$ supergigantic black holes.
Conventional wisdom is that the present entropy due
entirely to SMBHs is $\sim 10^{-8}$ of 
this asymptopic value. 
DMBHs increase the fraction 
up to $\sim 10^{-7}$, closer to asymptopia
and therefore more probable according to the second law of thermodynamics.
Note that a present fraction greater than $\sim 10^{-7}$
is not possible consistent with the present
observational data.

\bigskip

There are several previous arguments\cite{Ostriker,Moore,Murali,Yoo}
about the existence of DMBHs and they have put upper limits 
on their fraction of the halo mass.
The entropy arguments are new and provide additional motivation
to tighten these upper bounds or discover the halo black holes.
One observational method
is high longevity microlensing events. It is up to the
ingenuity of observers to identify other,
possibly more fruitful, methods some of which have
already been explored in a preliminary way.

\newpage

\begin{center}

\section*{Acknowledgements}

\end{center}

This work was supported in part 
by the U.S. Department of Energy under Grant
No. DG-FG02-06ER41418.

\end{document}